\documentclass[aps,superscriptadress,preprint]{revtex4}
\usepackage{graphicx}
\usepackage{amsmath}
\usepackage{mathrsfs}
\usepackage{amssymb}
\usepackage{booktabs}
\usepackage{tabularx}
\usepackage{textcomp}
\usepackage{feynmf}
\usepackage{color}
\begin{document}
\title{Two Flavor Baryon Number Susceptibilities of Non-equilibrium State}

\author{A-Meng Zhao$^{1}$}~\email[]{Email:zhaoameng@cxxy.seu.edu.cn}
\address{$^{1}$ Department of Foundation, Southeast University Chengxian College, Nanjing 210088, China}

\begin{abstract}
In this paper, to be closer to the realities of the relativistic heavy ion collisions than the thermal equilibrium assumption often used in the calculations of baryon number susceptibilities, the creatures of the collisions are treated as isotropic systems with temperature gradient. In addition, the quarks are thought as the quasi-particles with thermal masses. By doing these and according to Boltzmann transport equation, we get our results of the experimental observables $S\sigma$ and $\kappa\sigma^2$. We found that the results show a gap in value and a centrality dependence, which might indicate the relation between the baryon number susceptibilities and the temperature gradient in QGP.


\bigskip

\end{abstract}

\maketitle

\section{Introduction}
RHIC has undertaken its first phase BES(Beam Energy Scan) Program to search for the critical point and phase boundary in the QCD phase diagram\cite{b1}. Since the moments of the distributions of conserved quantities, for example net-baryon number, in the relativistic heavy ion collisions are related to the correlation length $\xi$ of the system\cite{b2}, they are believed to be good signatures of the QCD phase transition and its critical point. By relating the moments of the baryon number to the various order baryon-number susceptibilities\cite{b3}, the experimental observables $S\sigma$ and $\kappa\sigma^2$ could be calculated.

In the former works of our study on the baryon-number susceptibilities\cite{b4,b5,b6}, the quark-gluon plasma is often treated as a system of grand canonical ensemble. At the same time, the quark-gluon plasma created in the relativistic heavy ion collisions is considered to be undertaken the hydrodynamic evolution\cite{b7}. That is to say, the distributions of the quarks in the experiments deviate from those of thermal equilibrium state. Since the complexity of the quantum chromodynamics and the hydrodynamic evolution, we try to make a simplification to show the influence of the evolution on the the baryon-number susceptibilities qualitatively. To do this, we describe the quarks as quasi-particles with thermal masses\cite{b8,b9} and treat the creatures of the heavy ion collisions as isotropic systems with temperature gradient only. Though it can not describe the all realities in the heavy ion collisions, our attempt is closer to the realities comparing with the thermal equilibrium assumption. In addition, the relation between the susceptibilities and the factor of temperature gradient, among a set of factors, could be demonstrated more clearly.

\section{QGP with Temperature Gradient}
\label{two}

According to Boltzmann transport equation:
\begin{equation}
\frac{\partial{f}}{\partial{t}}+\frac{\partial{f}}{\partial{x}}\frac{dx}{dt}+\frac{\partial{f}}{\partial{y}}\frac{dy}{dt}+\frac{\partial{f}}{\partial{z}}\frac{dz}{dt}+\frac{\partial{f}}{\partial{v_x}}\frac{dv_x}{dt}+\frac{\partial{f}}{\partial{v_y}}\frac{dv_y}{dt}+\frac{\partial{f}}{\partial{v_z}}\frac{dv_z}{dt}=-\frac{f-f^0}{\tau_0},
\end{equation}
as we have mentioned above, if we take the temperature gradient only into account, then the equation is written as:
\begin{equation}
\frac{\partial{f}}{\partial{T}}\frac{\partial{T}}{\partial{x}}\frac{dx}{dt}+\frac{\partial{f}}{\partial{T}}\frac{\partial{T}}{\partial{y}}\frac{dy}{dt}+\frac{\partial{f}}{\partial{T}}\frac{\partial{T}}{\partial{z}}\frac{dz}{dt}=-\frac{f-f^0}{\tau_0},
\end{equation}
furthermore, if the distribution is isotropic, that is to say $\frac{\partial{T}}{\partial{x}}=\frac{\partial{T}}{\partial{y}}=\frac{\partial{T}}{\partial{z}}$ , we can get it as:
\begin{equation}
f=f^0-\alpha v_x\frac{\partial{f^0}}{\partial{T}}-\alpha v_y\frac{\partial{f^0}}{\partial{T}}-\alpha v_z\frac{\partial{f^0}}{\partial{T}},
\end{equation}
where $\alpha=\tau_0\frac{\partial{T}}{\partial{x}}$. Commonly, $\alpha$, as a function of time and position, is determined by the nature of QCD and the hydrodynamic evolution. Here we choose a constant $\bar{\alpha}$ instead of $\alpha$. And $\bar{\alpha}$ is the average of $\alpha$ over time and position.

In this paper, we adopt the quasi-particle model of QGP in which the interaction of quarks and gluons is treated as an effective mass term\cite{b9}. The effective mass of quark is made up of the rest mass and the thermal mass. The relation is expressed as follow:
\begin{equation}
m_q^2=m_{q0}^2+\sqrt{2}m_{q0}m_{th}+m_{th}^2,
\label{mq}
\end{equation}
where $m_{q0}$ is the rest mass of up or down quark, and the temperature dependent quark mass $m_{th}$ is
\begin{equation}
m_{th}^2(T)=\frac{g^2T^2}{18} N_{f},
\end{equation}
where $N_{f}$ is the number of quark flavors. Since only up and down quarks are considered here, $N_{f}=2$. And $g^2$ is related to the two-loop order running coupling constant,
\begin{equation}
g^2=\frac{24\pi}{(33-2N_{f})\ln\frac{T}{\Lambda_T}}(1-\frac{3(153-19N_{f})}{(33-2N_{f})^2}\frac{\ln(2\ln{\frac{T}{\Lambda_T}})}{\ln{\frac{T}{\Lambda_T}}}).
\end{equation}
As in \cite{b10}, we choose $m_{q0}=6.5$MeV and $\Lambda_T=122.5MeV$ in this paper.

Now the number density of the quarks with a single flavor and color is:
\begin{equation}
n=\frac{1}{4\pi^3}\int d^3p f=\frac{1}{4\pi^3}\int d^3p(f^0-3\bar{\alpha} v_x\frac{\partial{f^0}}{\partial{T}}).
\end{equation}
where $f^0=\frac{1}{1+e^{\beta(\epsilon-\mu)}}-\frac{1}{1+e^{\beta(\epsilon+\mu)}}$\cite{b4,b11} and $\epsilon=(p^2+m_q^2)^{1/2}$.

According to the simulations of the hydrodynamic evolution in heavy ion collisions\cite{b12}, we can estimate the value of $\bar{\alpha}$ approximately. For example, in the Au-Au collisions at $\sqrt{S_{NN}}=200GeV$, the temperature decreases from $0.35GeV$  to $0.05GeV$  in  $4-6fm$  at the beginning and the temperature gradient reduces to an insignificant value at about $\tau_0=10fm/c$ . That is to say, the average $\bar{\alpha}$  of the parameter  $\alpha=\tau_0\frac{\partial{T}}{\partial{x}}$ is about $0.7GeV/c$ in that case. In this paper, $\bar{\alpha}$ is varied from $0$ to $1GeV/c$ and from now on we simply use $\alpha$ to represent $\bar{\alpha}$.

It is mentioned in \cite{b6} that,while considering the statistic independence of the flavors and the confinement of the colors, the variance of baryon number density is
\begin{eqnarray}
\begin{split}
\sigma^2 &=<(n_{B}-<n_{B}>)^2>=\frac{N_fN_c^2}{3^2}<(n-<n>)^2>\\
&=N_f<(n-<n>)^2>=N_f\cdot\chi^{(2)}.
\end{split}
\end{eqnarray}
Similarly, we could get
\begin{eqnarray}
\begin{split}
S\sigma &=\frac{<(n_{B}-<n_{B}>)^3>}{<(n_{B}-<n_{B}>)^2>}=\frac{N_f<(n-<n>)^3>}{N_f<(n-<n>)^2>}\\
&=\frac{T^2\chi^{(3)}}{T\chi^{(2)}}=T\frac{\chi^{(3)}}{\chi^{(2)}},
\end{split}
\label{1}
\end{eqnarray}
and
\begin{eqnarray}
\begin{split}
\kappa\sigma^2 &=\frac{<(n_{B}-<n_{B}>)^4>-3<(n_{B}-<n_{B}>)^3>}{<(n_{B}-<n_{B}>)^2>}\\
&=\frac{<(n-<n>)^4>-3<(n-<n>)^3>}{<(n-<n>)^2>}\\
&=T^2\frac{\chi^{(4)}}{\chi^{(2)}}.
\label{2}
\end{split}
\end{eqnarray}
where $\frac{\partial n}{\partial \mu}=\cdot\chi^{(2)}$, $\frac{\partial^2 n}{\partial \mu^2}=\chi^{(3)}$ and $\frac{\partial^3 n}{\partial \mu^3}=\chi^{(4)}$.
.
\section{Results and Comparings with Experiments}
\label{three}
\begin{figure}
\centering
\includegraphics[width=1\linewidth]{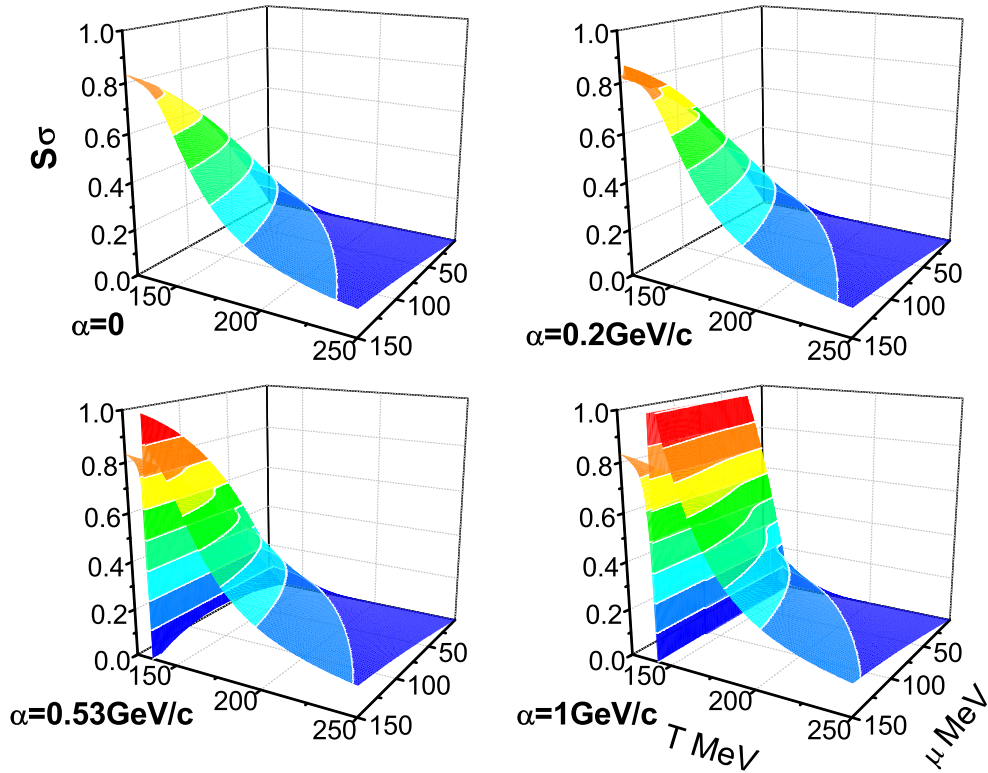}
\caption{$S\sigma$ is calculated at $\alpha=0,0.2,0.53$ and $1GeV/c$ respectively.}
\label{3}
\end{figure}

\begin{figure}
\centering
\includegraphics[width=1.0\linewidth]{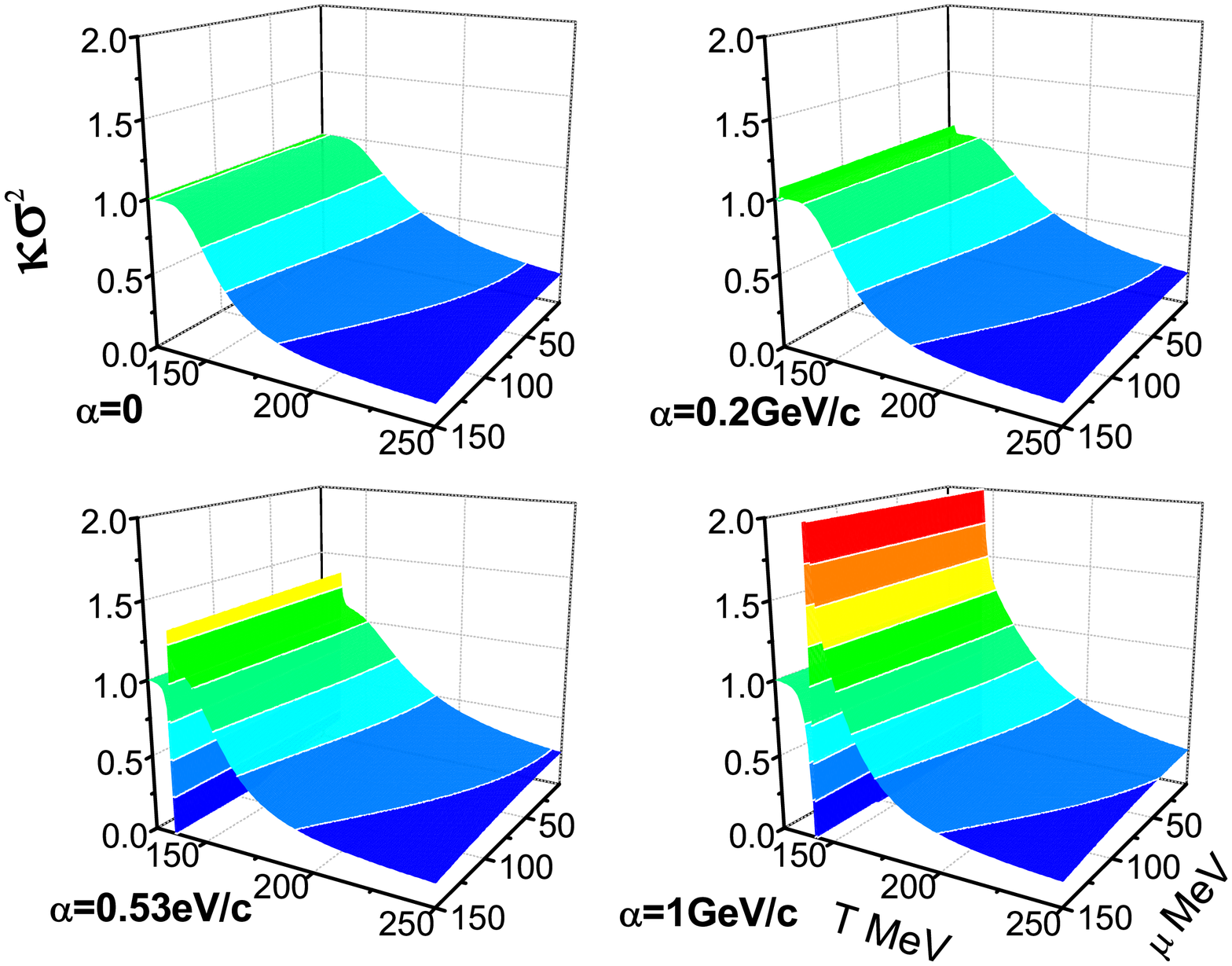}
\caption{$\kappa\sigma^2$ is calculated at $\alpha=0,0.2,0.53$ and $1GeV/c$ respectively.}
\label{4}
\end{figure}

Firstly, we shown the results of $S\sigma$ and $\kappa\sigma^2$ as a function of the temperature $T$ and the chemical potential of quark $\mu$. By Eq.(\ref{1}) and Eq.(\ref{2}), $S\sigma$ and $\kappa\sigma^2$ are calculated at the parameter $\alpha=0,0.2,0.53$ and $1GeV/c$ respectively and the results are shown in Fig.\ref{3} and Fig.\ref{4}. When $\alpha=0$, both $S\sigma$ and $\kappa\sigma^2$ show a continuous change with $T$ and $\mu$. Meanwhile, the gap appears at $\alpha=0.2$ and increases at $\alpha=0.53$ and $1GeV/c$ . Now it might be concluded that the transport effect caused by the temperature gradient in QGP is not neglectable in the study of $S\sigma$ and $\kappa\sigma^2$.

\begin{figure}
\centering
\includegraphics[width=1.0\linewidth]{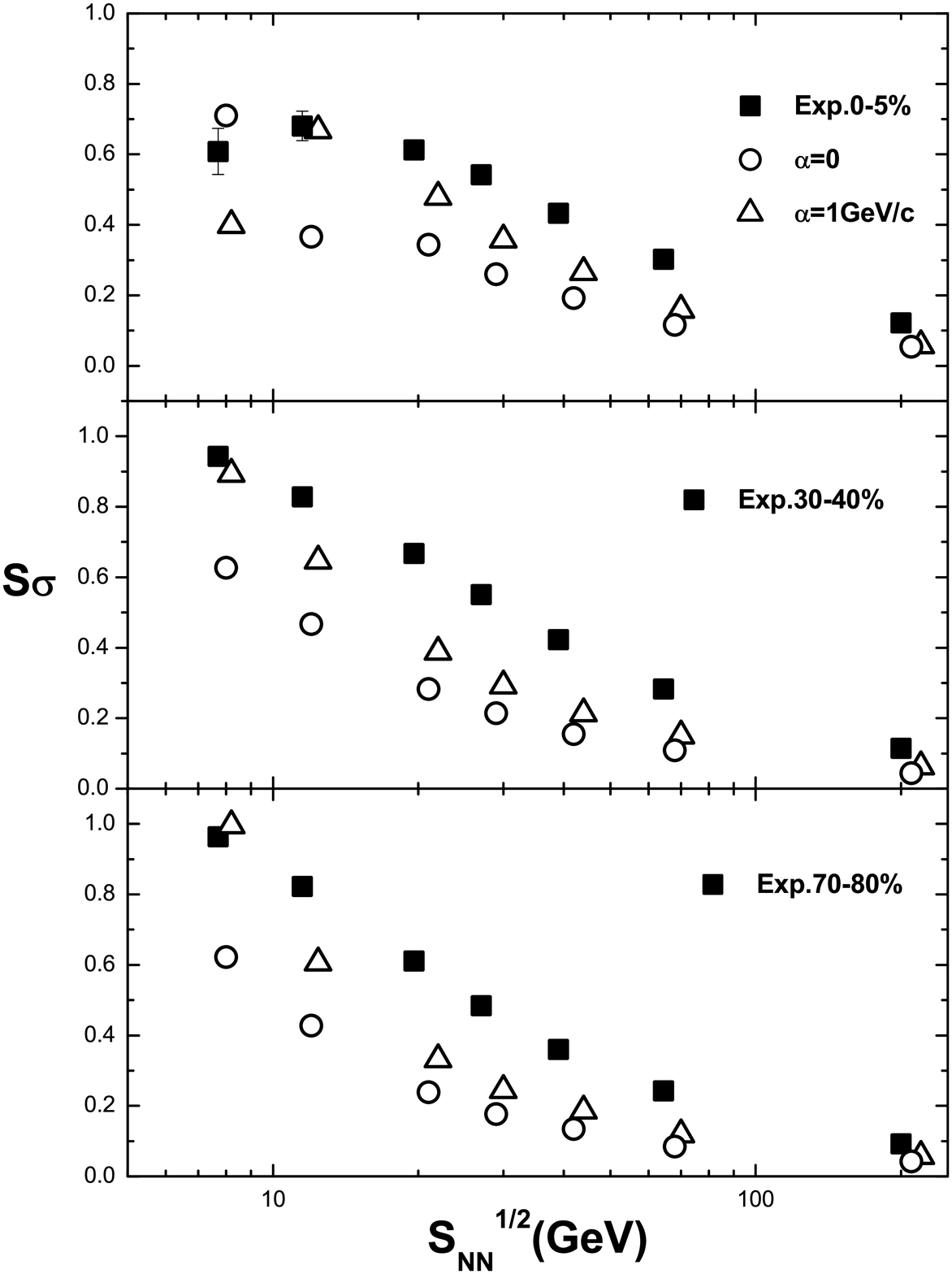}
\caption{$S\sigma$ of the quasi-particle model and the experiments. And experimental data comes from Ref.\cite{b13}.}
\label{5}
\end{figure}

\begin{figure}
\centering
\includegraphics[width=1.0\linewidth]{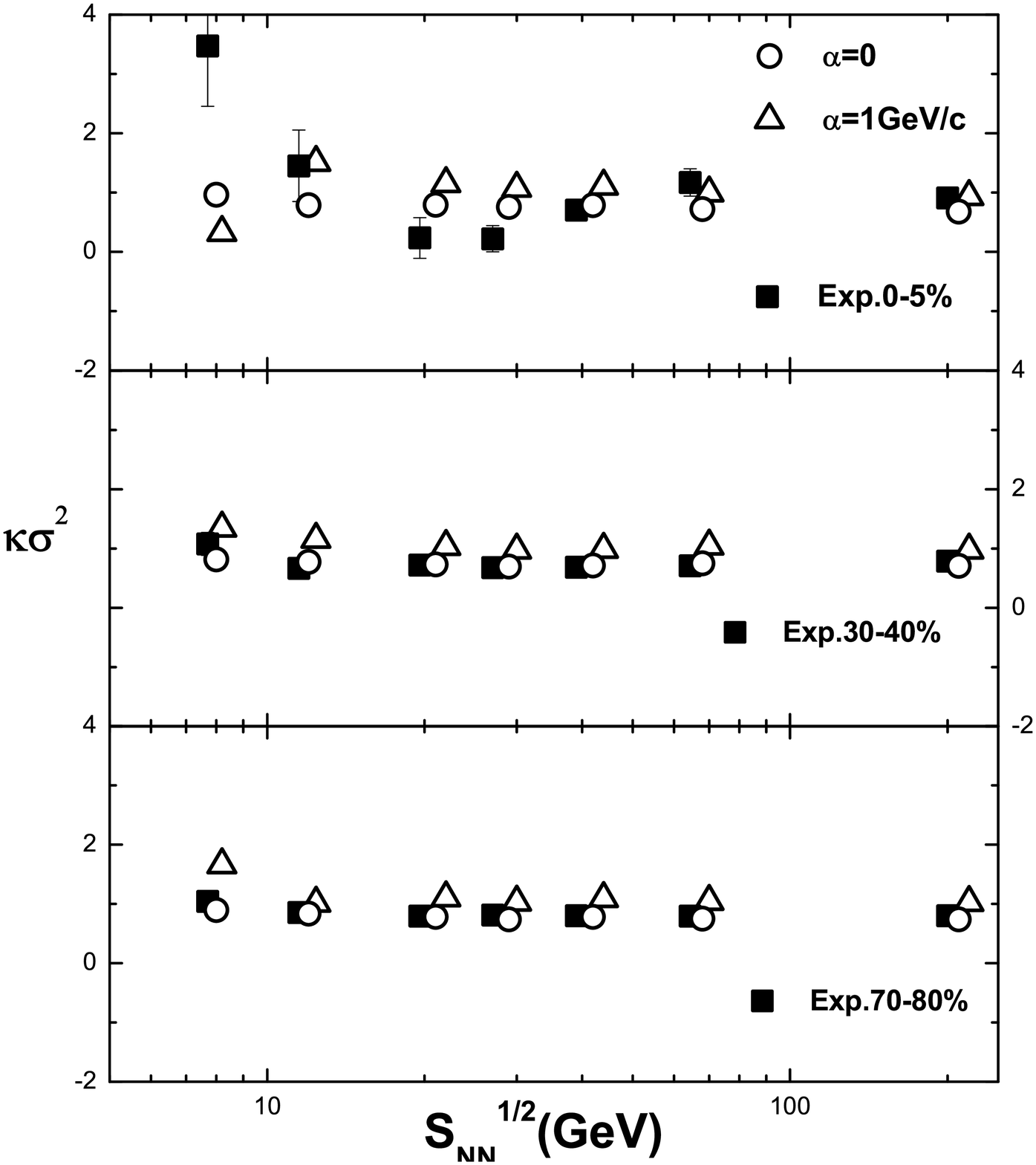}
\caption{$\kappa\sigma^2$ of the quasi-particle model and the experiments. And experimental data comes from Ref.\cite{b13}.}
\label{6}
\end{figure}

To test the reasonableness of our calculations, we compare the results of $S\sigma$ and $\kappa\sigma^2$ with the experimental observables in RHIC. The $S\sigma$ and $\kappa\sigma^2$  for net-proton distributions as a function of $\sqrt{S_{NN}}$ at different centralities of Au+Au collisions measured by STAR\cite{b13} are shown in Fig.\ref{5} and Fig.\ref{6}. The correspondence of the freeze-out $T$ and $\mu$ in the calculation of quasi-particle model to the collision energies $\sqrt{S_{NN}}$ comes from the experimental results\cite{b14}. At both $\alpha=0$ and $\alpha=1GeV/c$, the $S\sigma$ and $\kappa\sigma^2$  results are comparable to the experimental data, which  shows the reasonableness of the quasi-particle model we adopted.

At the same time, it should be noticed that the results in RHIC of these observables show not only an energy but also a centrality dependence\cite{b15}. As for our results at $\alpha=0$ , the dependence on the centrality could hardly be seen. Meanwhile at $\alpha=1GeV/c$, the centrality dependence is obvious. It possibly indicates that the transport effects in the relativistic heavy ion collisions might be related to the centrality dependence of $S\sigma$ and $\kappa\sigma^2$.

\section{Summary}
\label{four}
Baryon number fluctuations are believed to be good signatures of the QCD phase transition and its critical point. And the fluctuations can be directly connected to the susceptibilities of the system computed in theoretical calculations. In this paper, be different from a system of the grand canonical ensemble, we treat the creatures of the heavy ion collisions as an isotropic system with temperature gradient only. By doing this, our calculations are closer to the realities comparing with the thermal equilibrium assumption. In addition, the relation between the susceptibilities and the factor of temperature gradient, among a set of factors, could be demonstrated more clearly. And for simplicity, we treat the QGP as the quasi-particles with a temperature related mass term.

According to our results of $S\sigma$ and $\kappa\sigma^2$, there are two point should be noticed. Firstly, with the increase of $\alpha$, the results of $S\sigma$ and $\kappa\sigma^2$  as a function of $T$ and $\mu$ get the gaps. Secondly, comparing with those at $\alpha=0$, the results at $\alpha=1GeV/c$ show obvious centrality dependence of the collisions. Now it might be concluded that the thermal transport effect is related to the results of baryon number susceptibilities in the relativistic heavy ion collisions.

For the further study of baryon number fluctuations, the hydrodynamic evolution and more complicated transport effects in QGP should be taken into account. And the exploration of the relation between the local thermalization and the phase transition of QGP is of the same importance.

\section*{Acknowledgements}

This work is supported by the University Natural Science Foundation of JiangSu Province China (under Grants No.17KJB140003 ).

\end{document}